\documentclass{article}
\usepackage{dcase2022,amsmath,graphicx,url,times,booktabs, tabularx, xcolor}
\usepackage{lipsum,subcaption}

\usepackage{graphicx}
\usepackage{caption}

\usepackage{booktabs} 
\usepackage{adjustbox} 
\usepackage{array} 
\usepackage{amssymb}

\usepackage{booktabs} 

\usepackage{footmisc}  
\usepackage{hyperref}
\usepackage{xurl}      

\title{FALL-E: A Foley Sound Synthesis Model and Strategies}

\name{Minsung Kang$\sthanks{Equal contribution.}$, Sangshin Oh$^{*}$, Hyeongi Moon, Kyungyun Lee, Ben Sangbae Chon
      }
\address{Gaudio Lab, Inc., Seoul, South Korea. \texttt{bc@gaudiolab.com}
 }

\begin{document}

\maketitle

\begin{sloppy}

\begin{abstract}
This paper introduces FALL-E, a foley synthesis system and its training/inference strategies. The FALL-E model employs a cascaded approach comprising low-resolution spectrogram generation, spectrogram super-resolution, and a vocoder. We trained every sound-related model from scratch using our extensive datasets, and utilized a pre-trained language model. We conditioned the model with dataset-specific texts, enabling it to learn sound quality and recording environment based on text input. Moreover, we leveraged external language models to improve text descriptions of our datasets and performed prompt engineering for quality, coherence, and diversity. 
FALL-E was evaluated by an objective measure as well as listening tests in the DCASE 2023 challenge Task 7. The submission achieved the second place on average, while achieving the best score for diversity, second place for audio quality, and third place for class fitness.

\end{abstract}


\section{Introduction}
\label{sec:intro}
Generative AI has seen significant progress in recent years, particularly in the domains of images and text. However, the progress in sound generation has been comparatively slower. 

In the field of sound generation, numerous impressive works have been introduced including text-to-sound models such as AudioGen \cite{kreuk2022audiogen} and AudioLDM \cite{liu2023audioldm}. In addition, several works can be used as modules of the whole system such as Hifi-GAN \cite{hifigan}, SoundStream, EnCodec \cite{zeghidour2021soundstream, defossez2022high}, latent diffusion \cite{rombach2022high}, and spectrogram super-resolution \cite{sheng2019high}.

Furthermore, in text-input and text-conditioned generation, models such as T5 \cite{raffel2020exploring}, GPT \cite{radford2019language, brown2020language}, text prompt engineering \cite{liu2022design, white2023prompt}, and diffusion with conditioned generative models \cite{kreuk2022audiogen, liu2023audioldm, ramesh2021zeroshot, halgren2004glide} have been introduced. As the behavior of large deep learning models is somewhat difficult to analyze, these works enable us as users to steer the model using carefully selected text inputs.

In this context, we present a novel approach to foley synthesis that utilizes a cascade system composed of low-resolution spectrogram generation, a super-resolution module, and a vocoder. Our system represents our submission to the DCASE 2023 Task 7 - Foley Synthesis Challenge (Track A)~\cite{Choi_arXiv2023_01}. While we report objective measures with respect to the official evaluation set, our ultimate goal is to develop sound generation models that extend beyond the challenge's scope.

In Section 2, we introduce our model architecture, FALL-E, detailing the function of each module and how they work in tandem. In Section 3, we provide an in-depth analysis of our evaluation results, showcasing the effectiveness of our approach in various settings. Lastly, in Section 4, we summarize our contributions and highlight future directions for our work. 

\section{FALL-E}

\subsection{Architecture}

\begin{figure*}[t]
    \centering
    \includegraphics[width=\textwidth]{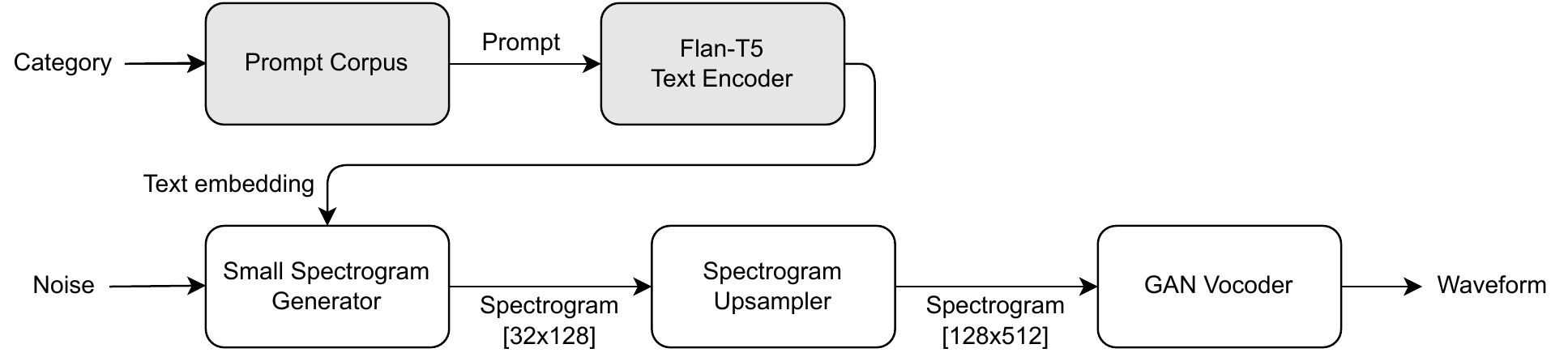}
    \caption{The overall system. Shaded blocks indicate the rule-based or pretrained models.}
    \label{fig:overall-system}
\end{figure*}

The cascade system, which involves generating low-resolution images or features and subsequently obtaining higher-resolution results, has been extensively utilized in generation models\cite{halgren2004glide, ho2022cascaded, ramesh2022hierarchical}. We adopt this approach to generate foley sound. Our proposed system, FALL-E, consists of three separately-trained models: diffusion-based low-resolution spectrogram generation model and upsampling model, and a GAN-based mel-spectrogram inversion network.

\textbf{Text Encoder} of FALL-E is a pre-trained Flan-T5, an instruction finetuned-variant of a T5 model which shows better performance for various applications \cite{chung2022scaling}. The class category is mapped to predefined text prompts from the prompt corpus. Then Flan-T5 converts the text prompts into a sequence of text embedding, which is input to the Low-resolution Spectrogram Generator.  

\textbf{Low-resolution Spectrogram Generator}\ is based on Glide, a diffusion generative model for text-to-image generation \cite{halgren2004glide}.  This module produces a low-resolution spectrogram. Specifically, it generates a 32 $\times$ 128 feature map for a 128-bin, 512-frame mel-spectrogram. The module employs a U-Net shaped architecture with 5 residual blocks in both the encoder and decoder. In the encoder, each block comprises 2 convolution layers and an additional upsampling layer with the number of convolution channels in each block increasing linearly from 192. The decoder is a mirrored version of the encoder.

\begin{table}[t]
\centering
\begin{tabular}{l r}
\toprule
\textbf{Module name}          & \textbf{Num. of Parameters} \\ \hline 
Text Encoder           & 110 M \\
Low-res. Spec. Generator  & 318 M \\
Spectrogram Upsampler  &  89 M \\
Mel Inversion Network  & 125 M \\
\midrule
Total                  & 642 M \\
\bottomrule
\end{tabular}%
\caption{
The number of parameters in each module.
}
\label{tab:num_params}
\end{table}

\textbf{Spectrogram Upsampler} is another diffusion-based generative model that synthesizes mel-spectrograms from a given low-resolution spectrogram. The overall architecture of this model is a U-Net that is similar to  Low-resolution Spectrogram Generator but with a different number of blocks and channels.
Its encoder and decoder consists of 4 blocks and the number of convolution channel in the first block is 128.
Unlike Low-resolution Spectrogram Generator, it isn't conditioned on text; it is only conditioned by the low-resolution mel-spectrogram feature.

\textbf{Mel Inversion Network} converts the generated mel-spectrograms into waveforms. Based on HiFi-GAN \cite{kong2020hifi} and BigVGAN \cite{lee2022bigvgan}, we add FiLM \cite{perez2018film} layers as a residual connection. The additional layer helps the model to preserve signal characteristics of the conditioned spectrogram and improves the phase reconstruction quality. We open-sourced this mel inversion network, GOMIN.\footnote{https://github.com/ryeoat3/gomin}

The whole system has 642M parameters in total. Its details are described in Table~\ref{tab:num_params}.


\subsection{Datasets} 
Training datasets include various sources across private and public audio datasets, including AudioSet \cite{gemmeke2017audio}, CLOTHO \cite{drossos2020clotho}, FreeToUseSounds.\footnote{https://www.freetousesounds.com/all-in-one-bundle/}, Sonniss,\footnote{https://sonniss.com/gameaudiogdc} WeSoundEffects,\footnote{https://wesoundeffects.com/we-sound-effects-bundle-2020}
and ODEON.\footnote{https://www.paramountmotion.com/odeon-sound-effects}
To prevent data imbalances or the potential risks of model misbehavior, samples with speech or musical contents are filtered out based on their metadata. After the filtering, we used 3,815 hours of audio signals for training.

\begin{table}[t]
\centering
\resizebox{\columnwidth}{!}{%
\begin{tabular}{l c r r c c c}
\toprule
\textbf{Name} & \textbf{AQ}
    & \multicolumn{2}{c}{\textbf{Dataset Size}}
    & \multicolumn{3}{c}{\textbf{Modality}} \\
\cmidrule(r){3-4} \cmidrule(l){5-7}
&  
& \multicolumn{1}{c}{\textbf{Dura.}} & \multicolumn{1}{c}{\textbf{N. Files}}
& \multicolumn{1}{c}{\textbf{Lb}} & \multicolumn{1}{c}{\textbf{Cp}} &\multicolumn{1}{c}{\textbf{Vd}} \\
\midrule
\textbf{Public dataset} \\
AudioSet 
    & \textit{noisy}    & 5420 \textit{h}   & 1,951,460   & \checkmark  &  & \checkmark \\

Clotho 
    & \textit{noisy}    & 37.0 \textit{h}   & 5,929 &   & \checkmark &\\
Free To Use Sounds 
    & \textit{noisy}    & 175.7 \textit{h}  & 6,370     &   & \checkmark &  \\
Sonnis Game Effects 
    & \textit{clean}    & 84.6 \textit{h}   & 5,049     &   & $\triangle$   & \\
WeSoundEffects 
    & \textit{clean}    & 12.0 \textit{h}   &   488     &   &   $\triangle$     &   \\
Odeon Sound Effects 
    & \textit{clean}    & 19.5 \textit{h}   & 4,420     &   & $\triangle$  &   \\

\midrule
\textbf{Private dataset} \\
Private dataset  
    & \textit{clean}    & 3829 \textit{h}  &  371,116    &   & $\triangle$  &   \\

\bottomrule
\end{tabular}%
}
\caption{
A list of audio datasets. AQ: audio quality, Dura.: duration, N. Files: number of files. Modality columns refer to the existence of labels, captions, and videos, respectively. 
\textit{Clean} recording: Audio is recorded in well-treated environments and mastered for professional content production. \textit{Noisy}: dataset contains environmental noises or interference signals.
$\triangle$: Textual information included, not necessarily captions.
This table is partially from \cite{kreuk2022audiogen} and \cite{wu2023large}
}
\label{tab:dataset}
\end{table}


\subsection{Prompting Strategy}\label{sub:train_infer_detail}



Text conditioning can be optimized or engineered to improve the model behavior. 
One of our focuses was to control the recording condition/environment of the generated signals so that the model can learn from crowd-sourced, noisy datasets (low recording SNR) as well, while being able to produce high-quality audio.
Among the datasets we used, AudioSet, Clotho, and Free To Use Soounds were "\textit{noisy}" dataset. We append a special token that indicates \textit{noisy dataset} to the text input during training. For the other datasets, we append \textit{clean dataset} token. The impact of this additional token will be discussed in Section \ref{sec:eval}.
We also clean the text label (i.e., text normalization) by dropping some stop words and numbers.


Our model is designed to process natural language text. When we directly use the sound class name as input, we have observed that the diversity of the generated sound is not as sufficient as that of real sound samples from the training dataset. On the other hand, by employing a variety of text prompts for each class, our model is capable of generating a more diverse range of sounds. For example, for footstep sound class, we can provide prompts such as:``\textit{clean} recording, footsteps on snow", ``\textit{clean} recording, footsteps, running", and ``\textit{clean} recording, footsteps in a large room".

\section{Evaluation and Analysis}
\label{sec:eval}

\begin{table}[]
\centering
\resizebox{\columnwidth}{!}{%
\begin{tabular}{l | r | r r r | r}
\toprule
\textbf{Sound class}  & \textbf{WAS $\uparrow$} & \textbf{Qual. $\uparrow$} & \textbf{Fit. $\uparrow$} & \textbf{Div. $\uparrow$}  & \textbf{FAD $\downarrow$} \\ \hline 
Dog bark             & 7.984  & 7.612  & 8.223  & 8.250  & 11.456  \\
Footstep             & 6.865  & 6.455  & 7.082  & 7.250  & 5.959  \\
Gun shot             & 7.255  & 6.814  & 7.573  & 7.500  & 3.021  \\
Keyboard             & 6.989  & 6.814  & 7.157  & 7.000  & 4.090  \\
Motor vehicle & 6.881  & 6.446  & 7.131  & 7.250  & 6.173  \\
Rain                 & 6.243  & 5.928  & 6.306  & 6.750  & 5.738  \\
Sneeze \& cough      & 6.553  & 6.528  & 6.606  & 6.500  & 2.340  \\
\midrule
Average              & 6.967  & 6.657  & 7.154  & 7.214  & 5.540  \\
\bottomrule
\end{tabular}%
}
\caption{DCASE 2023 task 7 official results across all sound classes. \textbf{WAS} indicates ``Weighted Average Score", \textbf{Qual.} refers to audio quality, \textbf{Fit.} to category fitness, and \textbf{Div.} to diversity within the class. 
}
\label{tab:scores}
\end{table}

\begin{table}[]
\centering
\begin{tabular}{l | r | r r r | r}
\toprule
\textbf{Model}  & \textbf{WAS $\uparrow$} & \textbf{Qual. $\uparrow$} & \textbf{Fit. $\uparrow$} & \textbf{Div. $\uparrow$}  & \textbf{FAD $\downarrow$} \\ \hline 
Surrey             & 7.886  & 7.546  & \textbf{8.419}  & 7.500  & \textbf{3.621}  \\
LINE               & 7.339  & 6.444  & 7.529  & \textbf{8.750}  & 3.679  \\
HEU                & 4.877  & 3.800  & 5.142  & 6.500  & 5.685  \\
Baseline           & 2.688  & 2.930  & 2.447  & \multicolumn{1}{c|}{-}  & 13.412  \\
\midrule
Ours               & \textbf{7.984}  & \textbf{7.612}  & 8.223  & 8.250  & 11.456  \\
\bottomrule
\end{tabular}%
\caption{Comparison of the official results for the ``Dog Bark" sound class in DCASE 2023 Task 7 with other submission models.
}
\label{tab:dog_bark}
\end{table}


In DCASE 2023 Task 7, our model achieved 2nd place in subjective scores and 3rd place in FAD scores, with a specific breakdown of 2nd place in Audio Quality, 3rd place in Category Fit, and 1st place in Diversity. Table \ref{tab:scores} presents the details of each sound class. In this section, we will delve deeply into the topics of objective and subjective evaluations.

The right column in Table \ref{tab:scores} presents FAD scores across all sound classes using the official evaluation repositories.\footnote{\label{foot:challenge-github}https://github.com/DCASE2023-Task7-Foley-Sound-Synthesis}
Our approach outperforms the baseline approach in all classes, with notable improvements observed in the rain and moving motor vehicle classes. Furthermore, the subjective quality is significantly improved by our model in all classes. It should be acknowledged that FAD scores may not be indicative of other important aspects of audio quality such as clarity, high-SNR, and high-frequency components. Also, as FAD measures similarity between a reference set and a test set, improvement beyond reference is mismeasured as a degradation, including quantization noise and codec noise.
As evidenced in Table \ref{tab:scores} and Table \ref{tab:dog_bark}, our performance in the ``Dog Bark" sound class received the worst score in FAD, while achieving the highest score in the Weighted Average Score (WAS).

Our model was developed to generate high-quality audio suitable for real-world scenarios using the environment and audio quality prefixes. Despite most of the audio samples in our training dataset exhibiting poor audio quality due to background noise, babble noise, wind noise, device noise, and codec distortion, we confirmed our model produces high-quality audio. As discussed in Section \ref{sub:train_infer_detail}, we controlled the audio sample quality by adding a special token as a prefix to the original text. Given that audio quality cannot be evaluated objectively, we conducted a informal listening test for the same text with both \textit{clean} and \textit{noisy} prefixes. Depending on the prefix used, we observed impressive improvements in sound quality across all sound classes. As illustrated in Figure \ref{fig:comparison_prefixes}, we can clearly observe that the use of the \textit{clean} prefix had a discernible impact on the audio quality, as indicated by the mel-spectrogram images. This type of model steering by prompting has been popular in other domains, and to our best knowledge, our work is the first work that successfully shows it in audio generation. 

\begin{figure}[t!]
    \centering
    \includegraphics[width=\columnwidth]{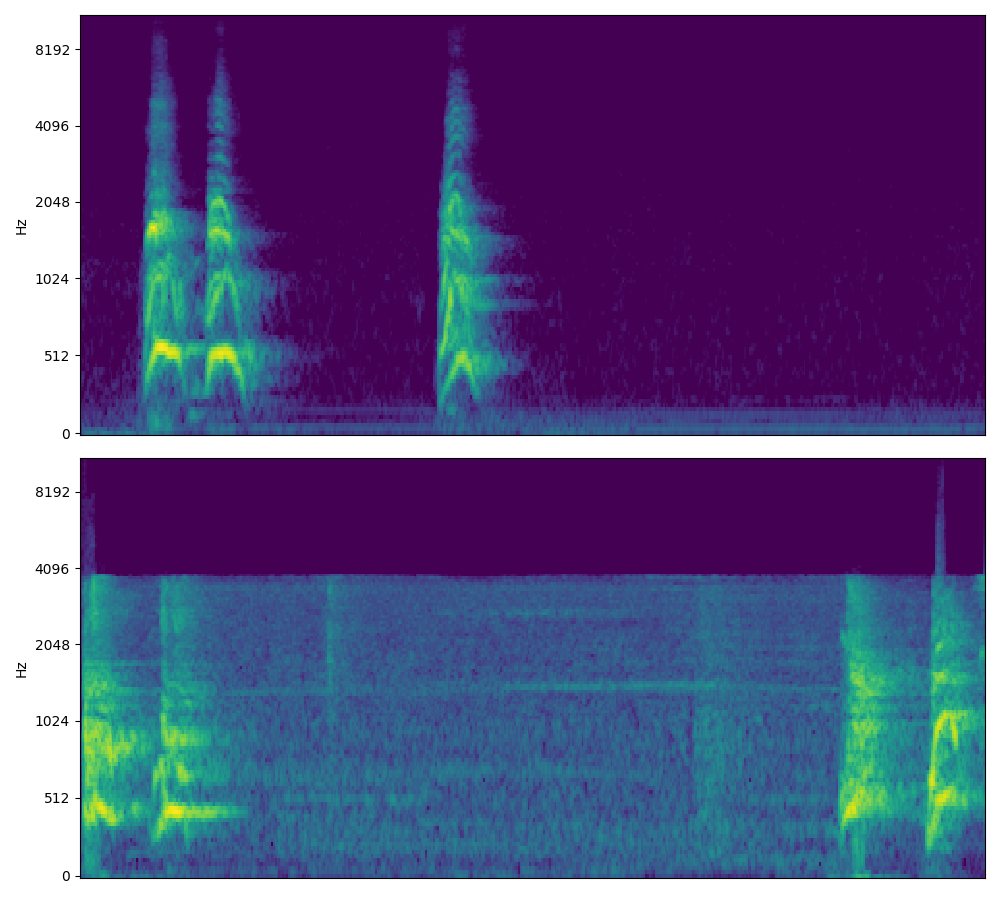}
    \caption{Mel-spectrograms of the generated audio samples using different recording environment prefixes. Prompts for images are (\textit{top}) ``\textit{clean} recording, puppy bark," and (\textit{botton}) ``\textit{noisy} recording, puppy bark," respectively}
    \label{fig:comparison_prefixes}
\end{figure}



\begin{figure}[t]
    \centering
    \includegraphics[width=\columnwidth]{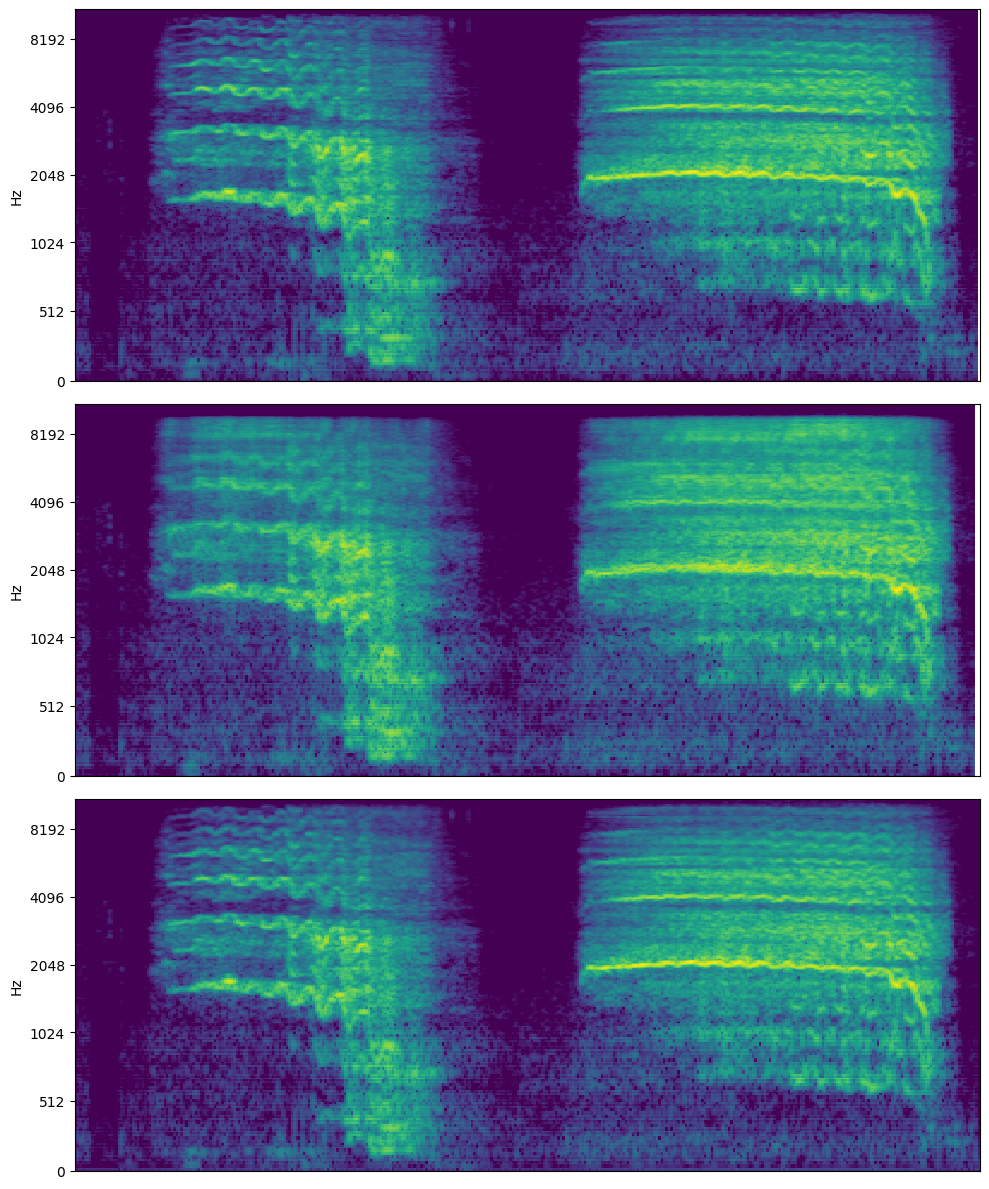}
    \caption{Mel-spectrogram for reconstructed samples. (\textit{top}) ground-truth waveform, (\textit{middle}) reconstructed with challenge baseline model, and (\textit{bottom}) Our GOMIN system.}
    \label{fig:comp-vocoder}
\end{figure}

To improve quality for mel-spectrogram inversion, we trained our own network based on HiFi-GAN \cite{kong2020hifi} and BigVGAN \cite{lee2022bigvgan} as explained above. Thanks to its bigger capacity and other architectural improvements, it showed better performance for overall sound categories. Compared to the baseline model,\footref{foot:challenge-github} our model well reconstructs tonal or harmonic components in the signal especially when the input mel-spectrograms include complex composition.

\section{Conclusion}
In this paper, we have presented FALL-E, Gaudio's foley synthesis system. FALL-E employs a cascaded approach with low-resolution spectrogram generation, a super-resolution module, and a vocoder. Our system was submitted to the DCASE 2023 Task 7 - Foley Synthesis Challenge (Track A), and we have reported the objective measure with respect to the official evaluation set. Through our extensive dataset and language model conditioning, as well as prompt engineering, we have achieved high-quality, diverse, and coherent sound generation results.

There is a vast potential for the development of generative AI in the audio domain. As technology continues to advance, new possibilities for sound generation arise, and the potential applications of this technology are vast. For example, in film and game production, foley synthesis could be used to produce more realistic sound effects, saving time and resources compared to traditional foley artistry. We believe that FALL-E, along with other works in the field, will pave the way for future advancements in generative audio technology, and we look forward to the continued development of this exciting area of research.

\section*{Acknowledgement}
This research was supported by Culture, Sports and Tourism R\&D Program through the Korea Creative Content Agency grant funded by the Ministry of Culture, Sports and Tourism in 2022 (Project Name: R\&D on AI Text-to-Sound Generation, Project Number: RS-2023-00229204, Contribution Rate: 100\%). The authors are grateful for Naver D2 Startup Factory and Naver Cloud Platform for supporting the GPU resources for this research.

We would like to highlight the clear arrangement implemented to ensure fairness and prevent any unfair advantage in the task. The conflict of interest with one of the organizers of this task was openly disclosed to the organizers, and the co-organizer affiliated with the institution in question remained uninvolved once the finalists were objectively determined. Additionally, during the subjective evaluation phase, other organizers were kept blind to the submission numbers to maintain impartiality. These measures were put in place to uphold the integrity and impartiality of the task evaluation process. 

\bibliographystyle{IEEEtran}
\bibliography{refs}

\end{sloppy}
\end{document}